\documentstyle[12pt,twoside,jltp,psfig]{article}
\newcommand{\beq}{\begin{equation}}
\newcommand{\eeq}{\end{equation}}

\title{Thermodynamics of a trapped Bose condensate \\
with negative scattering length} 

\author{B. Pozzi$^{(*)(+)}$, L. Salasnich$^{(*)(+)}$, 
A. Parola$^{(*)(++)}$, L. Reatto$^{(*)(+)}$ } 
\address{
$^{(*)}$ Istituto Nazionale per la Fisica della Materia, Unit\`a di Milano,\\ 
Via Celoria 16, 20133 Milano, Italy \\ 
$^{(+)}$ Dipartimento di Fisica, Universit\`a di Milano, \\ 
Via Celoria 16, 20133 Milano, Italy \\ 
$^{(++)}$ Dipartimento di Scienze Fisiche, Universit\`a dell'Insubria, \\ 
Via Lucini 3, 23100 Como, Italy} 

\begin{document}
\newpage

\begin{center}
{\bf Abstract}
\end{center}
\vskip 0.5 truecm

We study the Bose-Einstein condensation (BEC) for a system of $^7Li$ 
atoms, which have negative scattering length (attractive interaction), 
confined in a harmonic potential. Within the Bogoliubov and Popov 
approximations, we numerically calculate the density profile for both 
condensate and non-condensate fractions and the 
spectrum of elementary excitations. In particular, 
we analyze the temperature and number-of-boson dependence of these 
quantities and evaluate the BEC transition temperature $T_{BEC}$. 
We calculate the loss rate for inelastic two- and three-body 
collisions. We find that the total loss rate 
is strongly dependent on the density 
profile of the condensate, but this density profile does not 
appreciably change by increasing the thermal fraction. 
Moreover, we study, using the quasi-classical Popov approximation, 
the temperature dependence of 
the critical number $N_c$ of condensed bosons, for which there 
is the collapse of the condensate. 
There are different regimes as a function of the 
total number $N$ of atoms. For $N<N_c$ the condensate is always metastable 
but for $N>N_c$ the condensate is metastable only for 
temperatures that exceed a critical value $T_c$. 

PACS numbers: 03.75.Fi, 05.30.Jp, 32.80.Pj
\maketitle

\newpage
\section{INTRODUCTION}
\par
From 1995 we have experimental results interpreted as an 
evidence of Bose-Einstein condensation (BEC) 
in clouds of confined alkali-metal 
atoms \cite{b7}. In these dilute vapors one can separate off the 
effect of BEC and study the role of the interaction. 
For the theoretical understanding of these 
experiments, an important contribution comes from Pitaevskii 
and Gross with their idea of a space-time dependent 
macroscopic wavefunction (order parameter) that describes the 
Bose-condensed phase \cite{b8}. 
The experiments with alkali-metal atoms 
generally consist of three steps \cite{b9}:
a laser cooling and confinement in an external potential 
(a magnetic or magneto-optical trap), 
an evaporative cooling (the temperature 
reached is of the order of $100$ nK) and finally the analysis of 
the state of the system. 
Nowadays a dozen of experimental groups have achieved BEC by using 
different geometries of the confining trap and atomic species. 
\par 
In this paper we perform a study 
about the thermodynamics of a 
Bose-Einstein condensate of $^7$Li atoms,  
which is particularly interesting because of the 
attractive interatomic interaction: only a maximum 
number of atoms can form a condensate; beyond that the 
system collapses. Up to now 
there are only few experimental data supporting BEC 
for $^7$Li vapors with a limited condensate number. 
Because of the small number of condensed atoms, the size of the 
Bose condensate is of the same order of the resolution 
of the optical images. So it is difficult to obtain precise 
quantitative information, like the number of condensed atoms 
and the BEC transition temperature (see Bradley {\it et al.} \cite{b7}). 
A detailed theoretical study of the thermodynamical properties 
of such systems may give useful information for future experiments 
with a better optical resolution. 
\par
We consider an isotropic harmonic trap and concentrate our attention on the 
density profiles of condensate and non-condensate fraction and on 
the energy spectrum of elementary excitations by using both 
the Bogoliubov and Popov approximations. 
Actually, the Popov method and its quasi-classical approximation 
are a good starting point to estimate the contribution 
of the correlations in the gas for any finite value of the temperature 
\cite{b10}. 
\par
In Section II we describe the theoretical model used for our 
calculation and discuss in  detail the properties of systems with 
attractive interatomic interaction.
In Section III we illustrate the numerical procedure  
and the method we used to reduce numerical errors. Finally, 
in Section IV we analyze the numerical results. 

\section{THEORY} 
\par
The Heisenberg equation of motion for the bosonic field 
operator ${\hat \psi}({\bf r},t)$ of a non-relativistic 
system of confined and interacting 
identical atoms is given by \cite{b10,b11} 
\beq
i\hbar {\partial \over \partial t} {\hat \psi}({\bf r},t) =
\Big[ -{\hbar^2\over 2m} \nabla^2 
+ V_0({\bf r}) -\mu \Big] {\hat \psi} ({\bf r},t)
+ \int d^3{\bf r}' 
{\hat \psi}^+({\bf r}',t) V({\bf r},{\bf r}') 
{\hat \psi}({\bf r}',t){\hat \psi}({\bf r},t) \; ,
\label{e1}
\eeq
where $m$ is the mass of the atom, ${\mu}$ the chemical potential, 
$V_0$ is the confining external potential and 
$V$ is the interatomic potential. 
We separate out the condensate part setting  
\beq 
{\hat \psi}({\bf r},t)=\Phi({\bf r})+{\hat \phi}({\bf r},t) 
\label{e2}
\eeq
where 
$
\Phi({\bf r}) =\langle {\hat \psi}({\bf r},t) \rangle=\langle
{\hat \psi}({\bf r}) \rangle  
$
is the order parameter (macroscopic wavefunction) of the condensate, 
and $\langle ... \rangle $ is the mean value in the
grand-canonical ensemble. 
We can obtain an equation for $\Phi({\bf r})$ 
by inserting (\ref{e2}) into (\ref{e1}), by using the mean-field 
approximation for the fluctuations operators 
${\hat \phi}$ and ${\hat \phi}^+$ \cite{b10}, and by taking the mean value. 
To obtain the non-condensate fraction, we subtract from 
(\ref{e1}) its mean-value and use again the mean-field 
approximation. 
The fluctuation operator can be expanded in the following way
\beq 
{\hat \phi}({\bf r},t)= \sum_j \Big[ u_j({\bf r}) {\hat a}_j
e^{-iE_jt/\hbar} +v_j^*({\bf r}) {\hat a}_j^+
e^{iE_jt/\hbar} \Big] \; , 
\label{e6}
\eeq 
where ${\hat a}_j$ and ${\hat a}_j^+$ are bosonic operators and 
the complex functions $u_{j}({\bf r})$ and $v_{j}({\bf r})$ 
are the wavefunctions of the so-called 
quasi-particle excitations of energy $E_j$. 
\par
When the density of the atomic cloud is such that 
the scattering length and the range of the interatomic interaction 
are less than the average interatomic distance, 
the true interatomic potential can be 
approximated by a local pseudopotential 
$V({\bf r},{\bf r}')=g\delta^3({\bf r}-{\bf r}')$ \cite{b11}, 
where $g={4\pi \hbar^2 a_s/m}$ 
is the scattering amplitude of the spin triplet channel 
($a_s$ is the s-wave scattering length). 
In the typical conditions of confined $^7$Li vapors, 
the density is about $\rho=10^{20}$ m$^{-3}$ and 
$a_s=-1.4\cdot 10^{-9}$ m, i.e. $\rho |a_s|^3 <10^{-6}$, 
thus the pseudopotential gives reliable results. 
So we have the Hartree-Fock-Bogoliubov equation 
$$
{\hat L}u_j({\bf r})+gn({\bf r})
u_j({\bf r})+gm({\bf r})
v_j({\bf r})= E_j u_j({\bf r}) \; ,
$$
\beq
{\hat L}v_j({\bf r})+gn({\bf r})
v_j({\bf r})+gm({\bf r})
u_j({\bf r})= - E_j v_j({\bf r}) \; ,
\label{e7}
\eeq
where ${\hat L}=-{\hbar^2\over 2m} \nabla^2
+ V_0({\bf r}) -\mu
+ gn({\bf r})$,
and the $u_j({\bf r})$ and $v_j({\bf r})$ satisfy the
normalization condition
\beq
\int d^3{\bf r} \; [u_j^*({\bf r})u_k({\bf r})
- v_j^*({\bf r})v_k({\bf r})] = \delta_{jk} \; . 
\label{e7b}
\eeq
We write 
$n({\bf r})=\langle{\hat \psi}^+({\bf r}){\hat \psi}({\bf r})\rangle
=n_0({\bf r})+n_{nc}({\bf r})$ for the total local density,
$m({\bf r})=\langle{\hat \psi}({\bf r}){\hat \psi}({\bf r})
\rangle=
\Phi({\bf r}) \Phi^*({\bf r}) +m_{nc}({\bf r})$
for the total anomalous average function, 
$n_0({\bf r})=N_0|\Phi ({\bf r})|^2$ for the density 
of the condensate fraction ($N_0$ is the number of condensed atoms), 
$n_{nc}({\bf r})=\langle{\hat \phi}^+({\bf r}){\hat \phi}({\bf
r})\rangle$ for the local density of the non-condensate fraction, 
$m_{nc}({\bf r})=
\langle{\hat \phi}({\bf r}){\hat \phi}({\bf r}) \rangle$ 
for the anomalous average of the non-condensate fraction.
\par
Nevertheless, these equations remain very complex and 
some simplification is usually performed. 
\par 
One possible approximation, called Bogoliubov standard approximation, 
neglects the non-condensate fraction ($m_{nc}=n_{nc}=0$) in the equations. 
This approximation is particularly useful near zero temperature 
when the non-condensate is very small compared with the 
condensate fraction; so the condensate and 
its elementary excitations are separately calculated. 
Another possible, less drastic, simplification is 
called Popov approximation. 
It neglects only the anomalous average function $m_{nc}$ and it is able to 
give a more accurate description of the spectrum and 
non-condensate fraction 
(see Ref. 7 for the case of a homogeneous gas). 
Popov approximation is expected to be reliable in the whole range 
of temperature except near $T_{{}_{BEC}}$, where mean-field theories are 
known to fail. 
This set of equations must be solved with a self-consistent procedure 
that involves an iterative process. The equations are 
\beq
\Big[ -{\hbar^2\over 2m} \nabla^2
+ V_0({\bf r}) -\mu \Big] \Phi({\bf r})
+ g[n_0({\bf r})+2n_{nc}({\bf r})] \Phi({\bf r}) = 0 \; ,
\label{e10}
\eeq
$$
{\hat L}_P u_j({\bf r}) + g n_0({\bf r})v_j({\bf r})= E_j u_j({\bf r})
\; ,
$$
\beq
{\hat L}_P v_j({\bf r}) + g n_0({\bf r})u_j({\bf r})= - E_j v_j({\bf r})
\; ,
\label{e11}
\eeq
where ${\hat L}_P=-{\hbar^2\over 2m} \nabla^2 
+ V_0({\bf r}) -\mu + 2g n({\bf r})$. 
These equations are supplemented by the relation
fixing the total number of atoms in the system 
\beq
N=\int d^3 {\bf r} [n_0({\bf r})+n_{nc}({\bf r})] \; ,
\label{e12}
\eeq
where 
$$
n_0({\bf r})=N_0|\Phi ({\bf r})|^2
$$
\beq
n_{nc}({\bf r})=\sum_j \Big( |u_j({\bf r})|^2+|v_j({\bf r})|^2
\Big ) \langle{\hat a}^+_j {\hat a}_j\rangle+|v_j({\bf r})|^2
\label{e12b}
\eeq
with $\langle{\hat a}^+_j {\hat a}_j
\rangle=(e^{E_j/k_BT}-1)^{-1}$ the Bose factor ($k_B$ is the Boltzmann 
constant) at temperature $T$. 
The Popov approximation was first carried out in \cite{kohn} 
for repulsive interaction. 
\par
A possible way to avoid the solution of the partial differential 
equations (\ref{e10}) and (\ref{e11}) is to use the 
quasi-classical formula \cite{b13} for the Popov 
level of the excitation energies given by 
\beq
E({\bf p},{\bf r})=\sqrt{ \Big[ {p^2\over 2 m} + 
V_0({\bf r}) - \mu + 2 g n({\bf r}) \Big]^2 -g^2 n_0^2({\bf r}) } \; . 
\label{eqca} 
\eeq
With the help of this quasi-classical formula, 
the non-condensate density (thermal depletion) reads 
\beq
n_{nc}({\bf r}) = \int {d^3{\bf p}\over (2\pi \hbar)^3} 
\;\Big(-{\partial E \over \partial \mu}\Big)\;
\Big( e^{E({\bf p},{\bf r})/k_BT} - 1 \Big)^{-1} \; . 
\label{eqca1} 
\eeq
As suggested in Ref. 8, the quasi-classical Popov approximation 
is accurate only if $E >> E_0$, where $E_0$ 
is the lowest energy level of the elementary excitations. 
\par 
We study the BEC in an external harmonic potential 
with spherical symmetry, which is given by 
\beq 
V_{0}({\bf r})={1\over 2} m\omega^2 r^2 \; , 
\label{e13}
\eeq 
where $r=(x^2+y^2+z^2)^{1/2}$ and $\omega$ is the trap frequency. 
This symmetry makes easier the numerical procedure because 
one has to solve a one-dimensional problem. 
\par
For future reference, we recall some of the results in the case of 
non interacting particles (i.e. $a_s=0$) confined in an harmonic 
potential. 
In the limit of $k_{B}T>>\hbar\omega$ one 
can neglect the discretization of the eigenvalues 
(this corresponds to the limit 
$N\to\infty$ and $\omega\to0$ with $N\omega^3$ constant), 
and one gets the following result \cite{b15}:
\beq
T_{{}_{BEC}}^0={\hbar\omega \over k_B}\Big({N \over \zeta(3)}\Big)^{1/3}=
0.94{\hbar\omega \over k_B}N^{1/3} \; ,
\label{ei2}
\eeq
where $T_{{}_{BEC}}^0$ is the critical temperature, i.e. the temperature 
above which there is no a macroscopic population of the ground-state. 
The condensate fraction as a function of the temperature is given by 
\beq
{N_0 \over N}=1-\Big({T \over T_{{}_{BEC}}^0 }\Big)^3  \; . 
\label{ei3}
\eeq
The finite size correction to lowest order in $1/N$ 
of Eq. (\ref{ei2}) is \cite{b16} 
\beq
{\delta T_{{}_{BEC}}^0 \over 
T_{{}_{BEC}}^0}=-{\zeta(2) \over 2\zeta(3)^{2/3}}
N^{-1/3}
=-0.73N^{-1/3} \; .
\label{ei4}
\eeq
The finite size correction of Eq. (\ref{ei3}) is instead \cite{b15}:
\beq
{N_0 \over N}=1-\Big({T \over T_{{}_{BEC}}^0}\Big)^3-
{3\zeta(2) \over 2[\zeta(3)]^{2/3}}
\Big({T \over T_{{}_{BEC}}^0}\Big)^2N^{-1/3} .
\label{ei5}
\eeq
The correction in the condensate fraction due to the interaction 
has also been estimated as \cite{b13} 
\beq
{\delta T_{{}_{BEC}} \over T_{{}_{BEC}}^0}=-1.3{a_s \over a_h}N^{1/6} ,
\label{ei6}
\eeq
where $a_h=(\hbar / m\omega)^{1/2}$ is the characteristic length 
of the trap. We will compare the numerical solution of Popov equations 
with these analytical estimates.
Of course one should keep in mind that in a finite system 
no sharp transition can be present and, if present, it can only 
be an effect of approximation. 
\par 
In the case of negative scattering length, 
a simple variational calculation with a Gaussian trial wavefunction  
shows that a harmonic trap supports a Bose condensate at $T=0$ with 
a number of bosons smaller than a $N_c =0.67(a_h/|a_s|)$ \cite{sala1,sala2}) 
while numerical calculations give $N_c =0.57(a_h/|a_s|)$ \cite{b14}. 
In a homogeneous gas such critical number is zero. 
\par
For sake of completeness, we observe that, when a realistic nonlocal 
(finite range) interaction is taken into account, a new branch 
of Bose condensate appears for $^7$Li at higher density \cite{sala3,sala4}, 
but this non locality has a small effect on the low density branch 
unless the trap is much narrower of those of present experiments. 

\section{NUMERICAL PROCEDURE} 
\par 
To determine the order parameter of the condensate 
we use the steepest descent method \cite{b17}. 
The ground-state energy per particle and its wavefunction 
is obtained by minimizing the energy functional 
\beq
{\cal E}_p [\Phi ({\bf r})] = 
\int d^3{\bf r} {\Big [}{\hbar^2 \over 2m}|\nabla\Phi({\bf 
r})|^2+V_0({\bf r})|\Phi({\bf r})|^2+{gN \over 2}|\Phi({\bf r})|^4+
gn_{nc}({\bf r})|\Phi({\bf r})|^2{\Big ]} \; .
\label{e15}
\eeq 
Moreover, the ground-state wavefunction 
must satisfy the normalization condition $\int d^3{\bf r} 
|\Phi({\bf r})|^2=1$. 
After finite-difference spatial discretization\cite{sala5}, 
our function $\Phi(r)$ is represented by the vector ${\vec \Phi}$, 
whose elements are the values of $\Phi(r)$ on a grid of points 
of the radial axis. 
The functional (\ref{e15}) becomes a real function ${\cal E}({\vec \Phi})$. 
So, starting from a normalized trial vector 
${\vec \Phi}^{(0)}$, we can obtain the minimum by iterating the map 
\beq 
{\vec \Phi}^{(j+1)}={\vec \Phi}^{(j)}-
{\vec \nabla} {\cal E}({\vec \Phi}^{(j)}) \; ,
\label{e16} 
\eeq
and stopping the process when the change from the (j)-step 
to (j+1)-step is less than a fixed threshold. 
Note that the function must be normalized at each step. 
We use lattices up to $500$ points verifying that the 
results do not depend on the discretization parameters. 
\par 
Also the Bogoliubov and Popov elementary excitations 
are obtained by finite-difference discretization \cite{sala5} 
of Eq. (\ref{e11}) (to obtain the Bogoliubov levels we put 
$n_{nc}=0$ in these equations) and we 
have used a grid of $500$ points 
in the radial direction. 
Each eigenvector contains in its entries the discretized 
functions $u_j$ and $v_j$. In our case, 
each function takes up $500$ elements in a array of $1000$ 
so that the eigenvalue problem reduces to the 
diagonalization of a $1000\times 1000$ real matrix \cite{lapack}. 
\par 
It is necessary to perform other two operations to get our goal. 
First, we have to check the numerical errors 
in the evaluation of the eigenfunctions at the boundary of the mesh, 
where they must go to zero with a decreasing oscillatory behavior; 
second, we must impose the normalization condition (\ref{e7b}) 
and select eigenvalues and eigenvectors. 
We check that the sequence of maxima of each absolute value of 
the functions $u_j$ and $v_j$ is decreasing. When these 
functions are very close to zero, because of round off errors, 
this regular oscillatory and decreasing behavior is lost 
and we put the eigenfunctions equal to zero until 
the end of the grid. 
\par
For a given temperature $T$, 
we impose an energy threshold of $30\; k_BT$, 
and in any case 
in the sum (\ref{e12b}) are considered only eigenstates 
with energy smaller then a cut off ($70\%$ 
of the value of the potential in the last point of the grid). 
In this way we avoid to use the highest eigenvalues, those which are 
most sensitive to cut off errors. 
The symmetry of the system allows us to decompose 
the series (\ref{e12b}) and to consider separately 
the contribution of the levels with a fixed angular quantum 
number $l$. 
\par 
Our codes have been tested 
with solvable models, like the non-interacting case, 
by comparing numerical results with analytical solutions. 
We find that, for a good determination of the eigenvalues, 
it is not necessary a great number of points but one needs a 
large spatial range. A good choice, in our case, is a grid of $500$  
points in a range $0<r<35$ (in units of the characteristic 
length of the trap $a_h=3.23$ $\mu$m). 

\section{RESULTS}

\subsection{Gross-Pitaevskii equation} 

First of all we study the ground state properties by using the
equation (\ref{e10}) with $n_{nc}=0$, that is the 
Gross-Pitaevskii (GP) equation\cite{b8,gross}. 
This equation describes accurately 
the wave function of the condensate at zero temperature because 
the quantum depletion is negligible. 
We use the following numerical values for our system: 
m($^7$Li)$=6.941\cdot a.m.u.$, 
$\omega=878$ Hz for the frequency of the isotropic harmonic trap and 
$a_s=-27a_0$ for the scattering length\cite{lscatte} 
($a_0$ is the Bohr radius). 
We measure the radial coordinate 
$r$ in units of the characteristic oscillator length $a_h$ 
and the energies in units of energy oscillator quantum $\hbar \omega$. 
\par
The GP equation is solved for different values of the 
total number $N$ of atoms in the trap. 
The results are shown in Fig. 1, where we 
plot some properties of the solutions vs. $N$. All these curves 
show the existence of a critical threshold $N_c$. If we choose 
a $N>N_c$ we find that the solution of the GP equation is a sort of delta 
function centered at $r=0$ and the energy per particle goes 
to $-\infty$. The kinetic energy cannot 
compensate the increasing interaction energy any more: the energy 
functional has in these conditions only an absolute 
minimum that correspond to the collapsed system. 
Our numerical calculations show that the threshold is equal to 
$N_c=1287$, in full agreement with a previous numerical estimation 
\cite{b14}; instead, one finds $N_c=1514$ by using 
the approximate variational formula with a Gaussian trial wavefunction. 

\subsection{Far from the critical threshold $N_c$} 

Now we analyze in detail the system with $N=1000$, 
a number of atoms sufficiently far from the critical threshold 
$N_c$. We solve self-consistently the Popov equations (\ref{e10}) and
(\ref{e11}) for some value of the temperature. 
In this way, we obtain information about the density profiles 
of the condensate and non condensate fractions 
(\ref{e12b}) and the spectrum of elementary excitations. 
\par 
It is interesting to study the temperature 
dependence of the elementary excitations and the modification induced by the 
self-consistent Popov procedure. The changes are very small in both 
cases, as reported in Fig.2a and Tab.1. The energy spectrum is well 
represented by a linear function of $n$, i.e. by the spectrum 
of a harmonic oscillator with a small shift. 
A peculiar elementary excitation is that with radial quantum number $n=0$ 
and angular quantum number $l=1$. This is the dipole mode frequency 
at which the center of mass of the atomic cloud 
oscillates. Due to the generalized Kohn theorem 
such dipole mode should have the same frequency of the trap, 
but the Popov approximation does not fully respect this theorem 
because the thermal cloud is static \cite{kohn}. 
Except for the $n=1$ and $l=0$ mode, 
the Popov method gives eigenvalues lower than 
Bogoliubov ones. This effect is a consequence of the 
shape of the effective potential of the Bose condensate. In the Popov 
approximation, the effective potential is given by 
$V_0({\bf r})+2g(n_0({\bf r})+n_{nc}({\bf r}))$, thus it depends 
on the non condensate density profile too. 
This term gives, in the case of $l\not=$0, an effective potential 
with a larger parabolic shape and then with lower energy levels. 
The mode with quantum numbers $n=1$ and $l=0$, instead, is affected 
by the shape of the potential near $r=0$, where the term 
$2g(n_0({\bf r})+n_{nc}({\bf r}))$ produces a significant 
correction to the harmonic potential. 
So the presence of the non-condensed density profile gives a deeper 
well and a more bound state. 
In Fig. 2b we show the lowest energy levels as a function of the 
temperature. One sees that by increasing the temperature they 
approach even better 
the energy levels of the harmonic oscillator. 
This behaviour is due to the fact that by increasing the temperature 
the number of condensed atoms decreases and the nonlinear term 
of the eq. (\ref{e11}) is reduced. 
\par
In Fig. 3 the density profiles of the condensed and 
non condensed atoms are shown. 
We plot in Fig. 3a the profile of the condensate fraction for some 
value of temperature from $0$ to $58$nK. Note that 
the temperature does not modify the shape of the condensate density and 
its radial range. This is not the case of the non condensed density profile. 
As shown in Fig. 3b, the increasing temperature produces a  
sharper maximum at a larger value of $r$. 
Finally, in Fig. 4 we show the number of non condensed atoms 
for each value of the angular quantum number $l$, i.e. the population of the 
mode with a fixed quantum number $l$. For three values of the temperature, 
we compare the results in Bogoliubov and Popov approximation. 
The contribution for $l=0$ in Popov approximation is smaller 
than in Bogoliubov approximation because of the increase 
in the excitation energy obtained by the Popov 
method. We can see a change in the shape of 
$N_{nc}(l)$ with temperature, 
related to the modification of the non condensed 
density profiles. The more visible differences from 
Bogoliubov to Popov are due to the modes with lower quantum 
number $l$. Also this behavior is related to the form of the effective 
potential because the higher energy levels are not affected to the 
changes of the form of the potential as the lowest levels in the spectrum. 
\par
In Tab. 2 we compare the Bogoliubov and Popov results. 
The differences become more pronounced as temperature is increased. 
\par 
We summarize the results of our calculations with Popov method 
in Fig. 5, where we plot the condensate fraction $N_0/N$ versus the 
temperature $T$. In this figure we include also the 
quasi-classical approximation of the Popov and Hartree-Fock 
(i.e. to put the functions $v_j$ equal to zero) method 
and the curve of the ideal gas in the thermodynamic limit (\ref{ei2}). 
The critical temperature is in good agreement with the analytical 
estimation obtained by using (\ref{ei4}) and (\ref{ei6}): 
the correction (\ref{ei4}) shifts the critical temperature 
from $63$ to $58.4$ nK, whereas the correction (\ref{ei6}) 
from $58.4$ to $58.5$ nK. It is interesting 
to observe that the quasi-classical 
approximation of the Popov method gives accurate results 
for the condensate fraction at all temperatures here considered. 
To obtain the quasi-classical curve we have followed 
the self-consistent procedure described in Ref. 8, 
which avoids the numerical diagonalization 
of the Bogoliubov-Popov equations by using 
the quasi-classical formula (\ref{eqca}) for the excitation spectrum. 
It is also worth noting that 
in these conditions the Hartree-Fock approximation gives 
results compatible to those of the more accurate Popov method. 
\par
Finally, we evaluate the loss rate of the Bose condensate 
due to two- and three-body collisions. 
We use the relation \cite{loss}:
\beq 
\Gamma_L={dN_0 \over dt}=K \int d^3{\bf r} |n_0({\bf r})|^2 + 
L \int d^3{\bf r} |n_0({\bf r})|^3 
\label{loss1}
\eeq
where $K=1.2 \cdot 10^{-14} cm^3 s^{-1}$ is the coefficient of the 
two-body dipolar collisions and 
$L=2.6 \cdot 10^{-28} cm^6 s^{-1}$ is the coefficient 
of the three-body recombination collisions. 
These coefficients are strictly valid at zero temperature 
because they depend on temperature. Nevertheless, 
we can use Eq. (\ref{loss1}) to estimate the decay rate also 
at finite temperature because, in our self-consistent 
calculations, the condensate density is modified 
by the presence of the thermal cloud. We expect that 
our results are quite accurate near $T=0$. 
\par 
In Fig. 6 we plot, for $N=1000$, the mean radius of the condensate,  
obtained with the self-consistent Popov calculation, 
and the total loss rate $\Gamma_L$ as a function of temperature. 
By increasing the temperature, the number $N_0$ of condensed atoms 
decreases (see Fig. 5) and, because of the attractive interaction, 
the radius of the condensate grows: 
the loss rate is reduced due to a smaller probability of inelastic 
collision among condensed atoms. 
We can proceed in a different way by keeping fixed, at all $T$, 
the number $N_0$ of condensed atoms and, as a consequence, 
by increasing the total number $N$ of atoms in the trap. 
In this case the loss rate is practically constant 
because there is no significant modifications 
in the condensate density. 

\subsection{Near the critical threshold $N_c$} 

In this section we study the behavior of the Bose vapor 
by changing the number of the atoms. In particular we 
consider a total number of atoms near 
the critical threshold $N_c$, at which there is the 
collapse of the condensate. 
\par 
At zero temperature, by solving the Popov equations (\ref{e10}) 
and (\ref{e11}), 
we find $N_c=1260$, while the condensed atoms are $N_0^c=1256$. 
Remember that previously, by using the GP equation alone, 
i.e. neglecting the quantum depletion, we have found $N_c=N_0^c=1287$. 
This discrepancy is due to the properties of the spectrum 
of elementary excitations. By increasing $N$ there are no relevant changes 
in the the excitation energies besides the monopole mode 
$(n=1,\;l=0)$ that decreases. Strong oscillations 
begin when there is a level crossing between this mode and 
the dipole mode $(n=0,\;l=1)$ that has lower energy. 
At this point there are two modes with the same 
frequency and the system becomes unstable because of the 
induced resonance. 
\par 
We use the quasi-classical Popov approximation 
to evaluate the critical threshold $N_c$ as a function of 
the temperature. We modify the numerical 
procedure substituting the first Bogoliubov step by a Popov step, 
using a Boltzmann function $\exp{\big(-V_0({\bf r})/k_B T\big)}$ 
as trial non-condensed density. 
We have compared these quasi-classical results 
with the numerical ones of the full Popov approximation 
for some $T$ and $N$. 
For temperature less than about $80$ nK and total number of atoms 
less than about $4000$ we find a difference of about $3\%$ 
for the condensate fraction but 
when the temperature increases the quasi-classical method 
underestimates the condensed number of atoms. 
\par 
In Fig. 7a we plot the condensate fraction as a function of the 
temperature $T$ for different numbers of $^7$Li atoms. 
By increasing the temperature 
it is possible to put a larger number of atoms 
in the trap but the metastable condensate has a decreasing 
number of atoms. 
These results, obtained with the quasi-classical Popov approximation, 
show that, when $\omega= 878$ Hz, for $N<N_c=1260$ 
the system is always metastable and it has a finite 
condensate fraction for $0\le T<T_{{}_{BEC}}$. 
For $N>N_c$ the system is metastable only for 
temperatures that exceed a critical value $T_c$ and it has 
a finite condensate fraction for $T_c<T<T_{{}_{BEC}}$. 
Note that the critical number $N_0^c$ 
of condensed atoms slightly decreases by increasing the temperature. 
\par 
The collapse occurs because, by increasing the temperature, 
the effective potential $V_0({\bf r})+2gn_{nc}({\bf r})$ 
becomes tighter and tighter and so its characteristic length is reduced. 
We have seen that when the condensate fraction is only few percents 
then $n_{nc}(0)/n_0(0)$ becomes sufficiently large and 
the BEC cannot take place anymore. 
Thus, our calculations suggest that 
there should be a critical temperature $T^*$ 
beyond which the BEC transition is practically inhibited, 
independently of the number of atoms in the trap. 
The data in \cite{stoof} suggest that there is a very small 
fraction of condensed atoms also for temperature greater than $400$ nK. 
Actually, our rate of decrease of $N_c^0$ with $T$ is greater 
than the Hartree-Fock one obtained in Ref. 20.  
Note, however, that these mean-field methods are reliable when 
there is a large fraction of atoms in the condensate. 
\par
Obviously $T_c$ and $T_{{}_{BEC}}$ are functions of $N$. 
Such temperatures, and also the critical parameter $N_c$, 
are functions of the trap geometry, namely the frequency 
$\omega$ (see Fig. 7b), and of the scattering length $a_s$. 
\par 
Recently, the finite-temperature behavior 
of condensed $^7$Li near the critical threshold has been also studied 
by Davids {\it et al} \cite{22g}. They have used a different 
numberical scheme to solve self-consistently the Popov equations. 
Their results are consistent with ours: 
the critical number of condensed atoms 
slightly decreases by increasing the temperature. 
In particular, their calculations show that the rate of decrease 
of $N_0^c$ with $T$ is greater in the Popov treatment 
than in the Hartree-Fock treatment used by 
Houbiers and Stoof \cite{stoof}. 

\section{CONCLUSIONS} 

We have studied the thermodynamical properties of a 
weakly-interacting cloud of $^7$Li atoms confined in a harmonic trap. 
Due to the attractive interaction, the Bose condensed atoms 
remain in a metastable state until the number of atoms 
is lower than a critical threshold, that is about 
$10^3$ atoms at zero temperature. 
Far from the threshold, we have found that our 
Bogoliubov-Popov results give a transition temperature 
that is lower of about $5$ nK than the ideal gas value. 
This numerical value is in good agreement 
with the correcting formulas proposed in the literature \cite{b13,b16}. 
It is important to observe that, in these 
conditions, also the quasi-classical 
approximation of the Popov method gives a reliable estimation 
of the condensate fraction as a function of temperature. 
We have calculated the total loss rate (two- and three-body collisions) 
of the condensate: the loss rate is strongly influenced 
by the number of condensate atoms and by their density profile. 
This density profile is not affected by increased thermal 
fraction of atoms in the trap. 
\par 
Finally, we have studied the temperature dependence of 
the critical threshold. We have found that 
the critical number of condensed atoms decreases 
by increasing the temperature. 
The condensate presents strong oscillations 
when there is a crossing between the dipole mode and the 
monopole mode, that decreases 
for large numbers of atoms near the threshold. 
Such a behavior induces the collapse of the system: 
in this simple model, with local attractive interaction, 
the wavefunction of the condensate becomes a 
delta function and the energy per particle goes to minus infinity. 
We have shown that for $N<N_c$ the system is always metastable 
and it has a finite condensate fraction for $0\le T<T_{{}_{BEC}}$. 
Instead, for $N>N_c$ the system is metastable only for 
temperatures that exceeds a critical value $T_c$; in this case 
the system has a finite condensate fraction for $T_c<T<T_{{}_{BEC}}$. 
Moreover, within the quasi-classical Popov approximation, 
our calculations suggest that there should be a temperature $T^*$, 
beyond which the BEC transition is inhibited for any number of particles. 
\par
We believe that our calculations could be useful 
to select experimental conditions\cite{loro} 
for which a fraction of condensed $^7$Li atoms 
is detectable at thermal equilibrium. 

\section*{ACKNOWLEDGEMENTS}
\par
This work has been supported by INFM under the Research 
Advanced Project (PRA) on "Bose-Einstein Condensation". 


\newpage

\section*{Figure Captions} 

\vskip 1. truecm

{\bf Fig. 1}. Specific quantities of condensed $^7$Li atoms,  
obtained by solving the the Gross-Pitaevskii equation, 
as a function of the total number of atoms (N). 
The energies are in units of $\hbar \omega$, 
where $\omega=878$ Hz is the frequency of the external harmonic potential. 
The radius is in units of the characteristic length
$a_h=(\hbar / m\omega)^{1/2}=3.23$ $\mu$m of the trap.

\vskip 1. truecm

{\bf Fig. 2}. a) Elementary excitations 
of $10^3$ $^7$Li atoms at $T=58$ nK. 
Comparison of the Popov approximation 
with the Boboliubov one. Units as in Fig. 1. 
$n$ is the radial quantum number. 
b) The lowest Popov energy levels vs. temperature. 
$N=10^3$ and units as in Fig. 1. 

\vskip 1. truecm

{\bf Fig. 3}. Popov calculations for $10^3$ $^7$Li atoms. 
a) Profile of the condensate density for temperature ranging 
from $0$ to $58$nK. b) The same for the non-condensate density. 

\vskip 1. truecm

{\bf Fig. 4}. 
Number of non condensed atoms ($N_{nc}$) for each 
value of the angular quantum 
number $l$, with a comparison of the Popov 
approximation with the Bogoliubov 
one. The radius is in units of the characteristic length 
$a_h=(\hbar / m\omega)^{1/2}=3.23$ $\mu$m of the trap.  
 
\vskip 1. truecm 

{\bf Fig. 5}. Condensate fraction {\it vs} temperature for 
$10^3$ $^7$Li atoms. 

\vskip 1. truecm

{\bf Fig. 6}. Mean radius ($r_m$) of the condensate (top) 
and total loss rate (bottom) vs. the temperature for 
$10^3$ $^7$Li atoms. Units as in Fig. 1. 

\vskip 1. truecm 

{\bf Fig. 7}. Condensate fraction {\it vs} temperature for 
different numbers of $^7$Li atoms. Results obtained with 
the quasi-classical Popov approximation. Trap frequency: 
a) $\omega_1=878$ Hz; b) $\omega_2=439$ Hz. 

\newpage

\begin{center}
\begin{tabular} {||c||c|c|c|c||}
\hline \hline 
T (nK) & $(1,0)$ & $(0,1)$ & $(10,5)$ & $(10,10)$ \\
\hline 
 0 &	1.77 &  1.00 &  25.24 &  30.30 \\
10 &	1.77 &  1.00 &  25.23 &  30.29 \\
20 &	1.79 &  0.99 &  25.20 &  30.26 \\
30 &	1.83 &  0.99 &  25.13 &  30.18 \\
40 &	1.88 &  0.98 &  25.03 &  30.07 \\
50 &	1.92 &  0.96 &  24.91 &  29.93 \\
53 &	1.93 &  0.96 &  24.87 &  29.88 \\
58 & 	1.95 &  0.95 &  24.81 &  29.80 \\
58.5 &	1.95 &  0.95 &  24.80 &  29.79 \\
\hline \hline
\end{tabular}
\end{center}
{\bf Table 1}. Some excitation energies for $N=1000$ $^7$Li atoms 
with the Popov method at different values 
of the temperature $T$. The energies, labeled by 
quantum numbers $(n,l)$, are in units of $\hbar \omega$, 
where $\omega=878$ Hz is the frequency of the external harmonic potential. 

\newpage

\begin{center}
\begin{tabular} {||c||c|c||c|c||c|c||}
\hline \hline
T (nK) & $N_0^{(B)}$ & $N_0^{(P)}$ & $N_{nc}^{(B)}$ & $N_{nc}^{(P)}$ & 
$N_0^{(B)}/N$ & $N_0^{(P)}/N$ \\
\hline \hline
 0 &	999.58 &  999.58 &  0.42 &  0.42 &  99.96 &  99.96 \\
10 &	991.02 &  990.99 &  8.98 &  9.00 &  99.10 &  99.10 \\
20 &	950.07 &  949.52 &  49.93 &  50.48 &  95.01 &  94.95 \\
30 &	854.31 &  850.64 &  145.69 &  149.36 &  85.43 &  85.06 \\
40 &	680.75 &  667.18 &  319.24 &  332.82 &  68.07 &  66.72 \\
50 &	407.76 &  371.82 &  592.24 &  628.18 &  40.77 &  37.18 \\
53 &	303.47 &  257.48 &  696.52 &  742.52 &  30.35 &  25.75 \\
58 &	105.00 &  38.06 &  894.99 &  961.94 &  10.50 &  3.80 \\
58.5 &	83.41 &  13.97 &  916.59 &  986.02 &  8.34 &  1.40 \\
\hline \hline
\end{tabular}
\end{center}
{\bf Table 2}. Number of condensed ($N_0$) and non condensed ($N_{nc}$) 
atoms and condensate fraction $N_0/N$ for $N=1000$ $^7$Li atoms at 
different values of the temperature $T$. 
$^{(B)}$ denotes the Bogoliubov result and $^{(P)}$ the Popov one. 


\begin{thebibliography}{100}

\bibitem{b7} {M.H. Anderson, J.R. Ensher, M.R. Matthews, C.E. Wieman, 
and E.A. Cornell, {\it Science} {\bf 269}, 189 (1995); 
K.B. Davis, M.O. Mewes, M.R. Andrews, N.J. van Druten,
D.S. Drufee, D.M. Kurn, and W. Ketterle, {\it Phys. Rev. Lett.} {\bf 75},
3969 (1995); C.C. Bradley, C.A. Sackett, J.J. Tollett, and R.G. Hulet, 
{\it Phys. Rev. Lett.} {\bf 75}, 1687 (1995); 
C.C. Bradley, C.A. Sackett, and R.G. Hulet, 
{\it Phys. Rev. Lett.} {\bf 78}, 985 (1997).}

\bibitem{b8} {L.P. Pitaevskii, {\it Zh. Eksp. Teor. Fiz.} {\bf 40}, 646 (1961) 
[English. Transl. {\it Sov. Phys.} JETP {\bf 13}, 451 (1961)].}

\bibitem{b9} {W. Ketterle, M.R. Andrews, K.B. Davis, D.S. Drufee, 
D.M. Kurn, M.O. Mewes, and N.J. vanDruten, {\it Physica Scripta} {\bf T66}, 
31 (1996).}

\bibitem{kohn} {D.A.W. Hutchinson, E. Zaremba, and A. Griffin, 
{\it Phys. Rev. Lett.} {\bf 78}, 1842 (1997).}

\bibitem{b10} {A. Griffin, {\it Phys. Rev. B} {\bf 53}, 9341 (1996).} 

\bibitem{b11} {K. Huang, Statistical Mechanics (John Wiley, New York, 1963).}

\bibitem{gross} {E.P. Gross, {\it Nuovo Cimento} {\bf 20}, 454 (1961); 
{\it J. Math. Phys.} {\bf 2}, 195 (1963).}

\bibitem{b12} {V.N. Popov, Functional Integrals and Collective Modes 
(Cambridge University Press, New York, 1987), cap.6.} 

\bibitem{b13} {S. Giorgini, L.P. Pitaevskii, and S. Stringari, 
{\it Phys. Rev. A} {\bf 54}, 4633 (1996); 
{\it Phys. Rev. Lett.} {\bf 78}, 3987 (1997); 
{\it J. Low Temp. Phys.} {\bf 109}, 309 (1997).}

\bibitem{lscatte} {E.R.I. Abraham, W.I. McAlexsander, C.A. Sackett,
and R.G. Hulet, {\it Phys. Rev. Lett.} {\bf 74}, 1315 (1995).}

\bibitem{b14} {P.A. Ruprecht, M.J. Holland, K. Burnett, and M. Edwards, 
{\it Phys. Rev. A} {\bf 51}, 4704 (1995).}

\bibitem{b15} {V.V. Goldmann, I.F. Silvera, and A.J. Leggett, 
{\it Phys. Rev. B} {\bf 24}, 2870 (1981); V. Bagnato, D.E. Pritchard, 
D. Kleppner, {\it Phys. Rev. A} {\bf 35}, 4354 (1987).}

\bibitem{b16} {S. Grossman and M. Holthaus, {\it Phys. Lett. A} {\bf 208}, 188
(1995); K. Kirsten and D.J. Toms, {\it Phys. Rev. A} {\bf 54}, 4188 (1996); 
W. Ketterle and N.J. vanDruten, 
{\it Phys. Rev. A} {\bf 54}, 656 (1996); 
H. Haugerund, T. Haugset, and F. Ravndal, 
{\it Phys. Lett. A} {\bf 225}, 18 (1997).}

\bibitem{b17} {S. Koonin and C.D. Meredith, 
Computational Physics (Addison-Wesley, Reading, 1990).}

\bibitem{sala1} {L. Salasnich, {\it Mod. Phys. Lett. B} {\bf 11}, 1249 (1997); 
{\it Mod. Phys. Lett. B} {\bf 12}, 649 (1998).} 

\bibitem{sala2} {E. Cerboneschi, R. Mannella, E. Arimondo, and L. 
Salasnich, {\it Phys. Lett. A} {\bf 249}, 245 (1998).}

\bibitem{sala3} {A. Parola, L. Salasnich, and L. Reatto, 
{\it Phys. Rev. A} {\bf 57}, R3180 (1998).} 

\bibitem{sala4} {L. Reatto, A. Parola, and L. Salasnich, 
{\it J. Low Temp. Phys.} {\bf 113}, 195 (1998).} 

\bibitem{sala5} {L. Salasnich, A. Parola, and L. Reatto, 
{\it Phys. Rev. A} {\bf 59}, 2990 (1999).}

\bibitem{lapack} {DGEEV subroutine, the Linear Algebra Package (LAPACK) 
driver routine.} 

\bibitem{stoof} {M. Houbiers and H.T.C. Stoof, {\it Phys. Rev. A} 
{\bf 54}, 5055 (1996).} 

\bibitem{loro} {C.A. Sackett, J.M. Gerton, M. Welling, and R.G. Hulet, 
{\it Phys. Rev. Lett.} {\bf 82}, 876 (1999).} 

\bibitem{loss} {R.J. Dodd, M.Edwards, C.J. Williams, and C.W. Clark, 
{\it Phys. Rev. A} {\bf 54 }, 661 (1996); H. Shi, W-M. Zheng, 
{\it Phys. Rev. A} 
{\bf 55}, 2930 (1997); C. Huepe, S. Metens, G. Dewel, P. Borckmans, 
and M.E. Brachet, {\it Phys. Rev. Lett.} {\bf 82}, 1616 (1999).}

\bibitem{22g} {M.J. Davis, D.A.W. Hutchinson, and E. Zaremba, 
Preprint cond-mat/9906334.} 

\end{thebibliography}
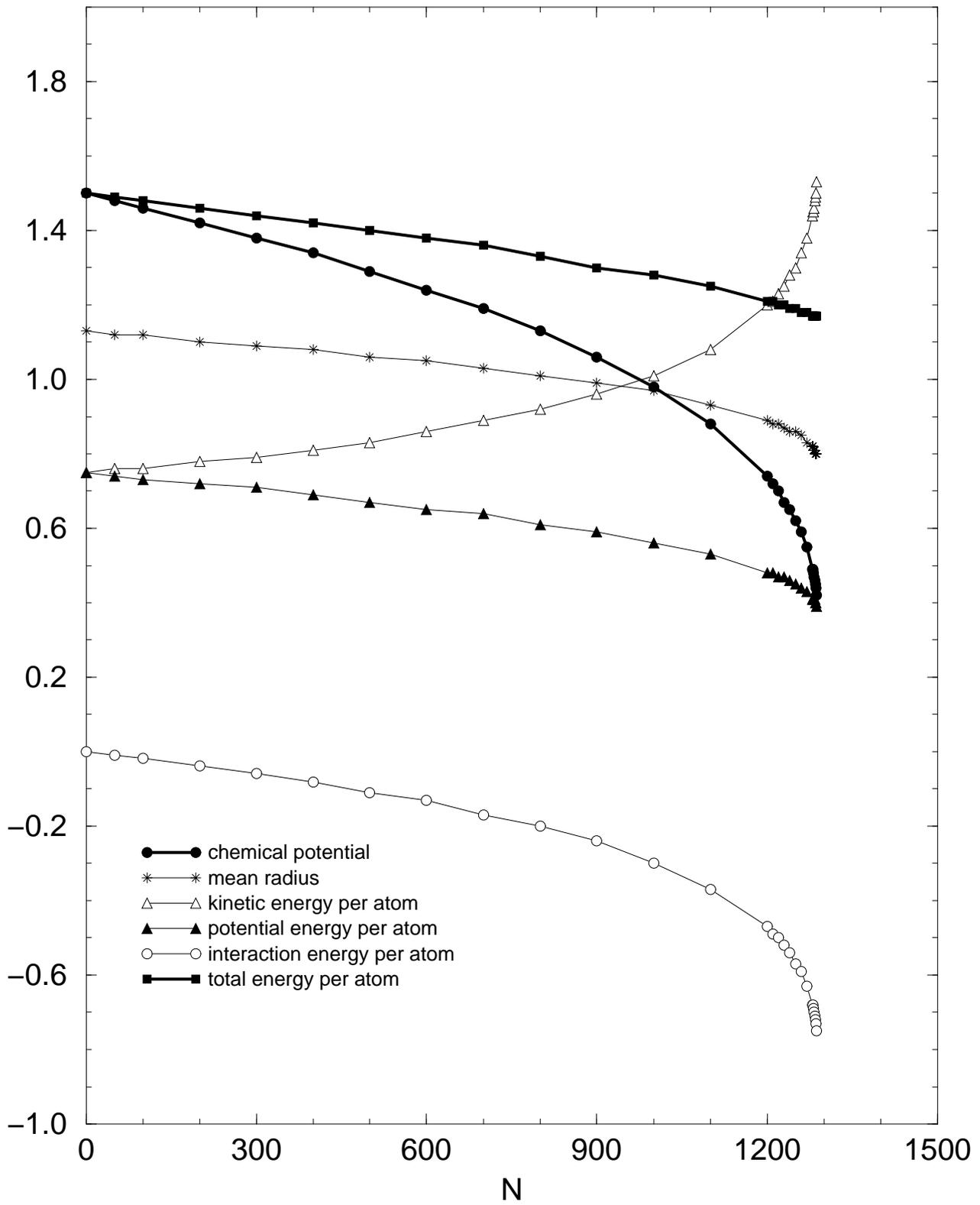
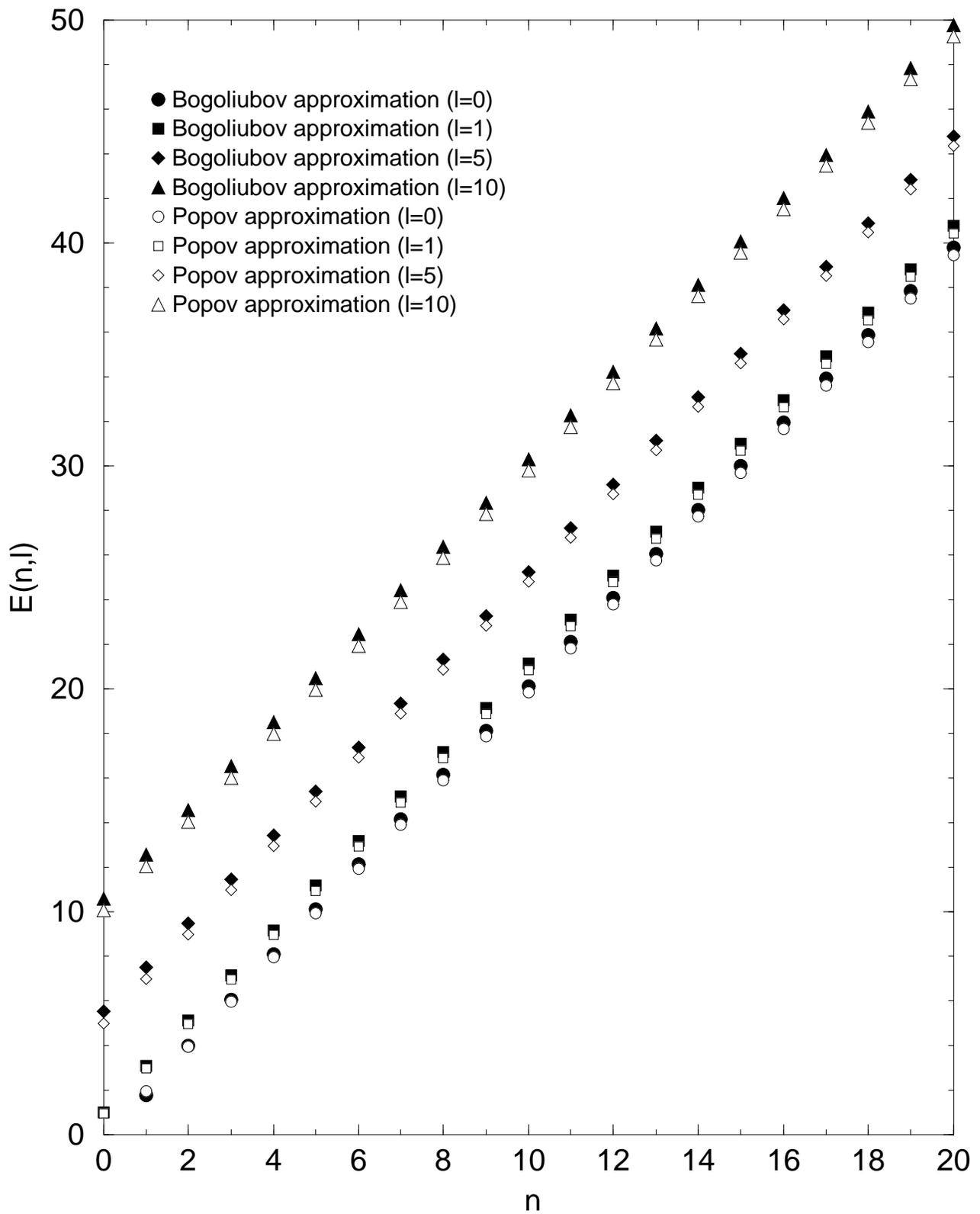
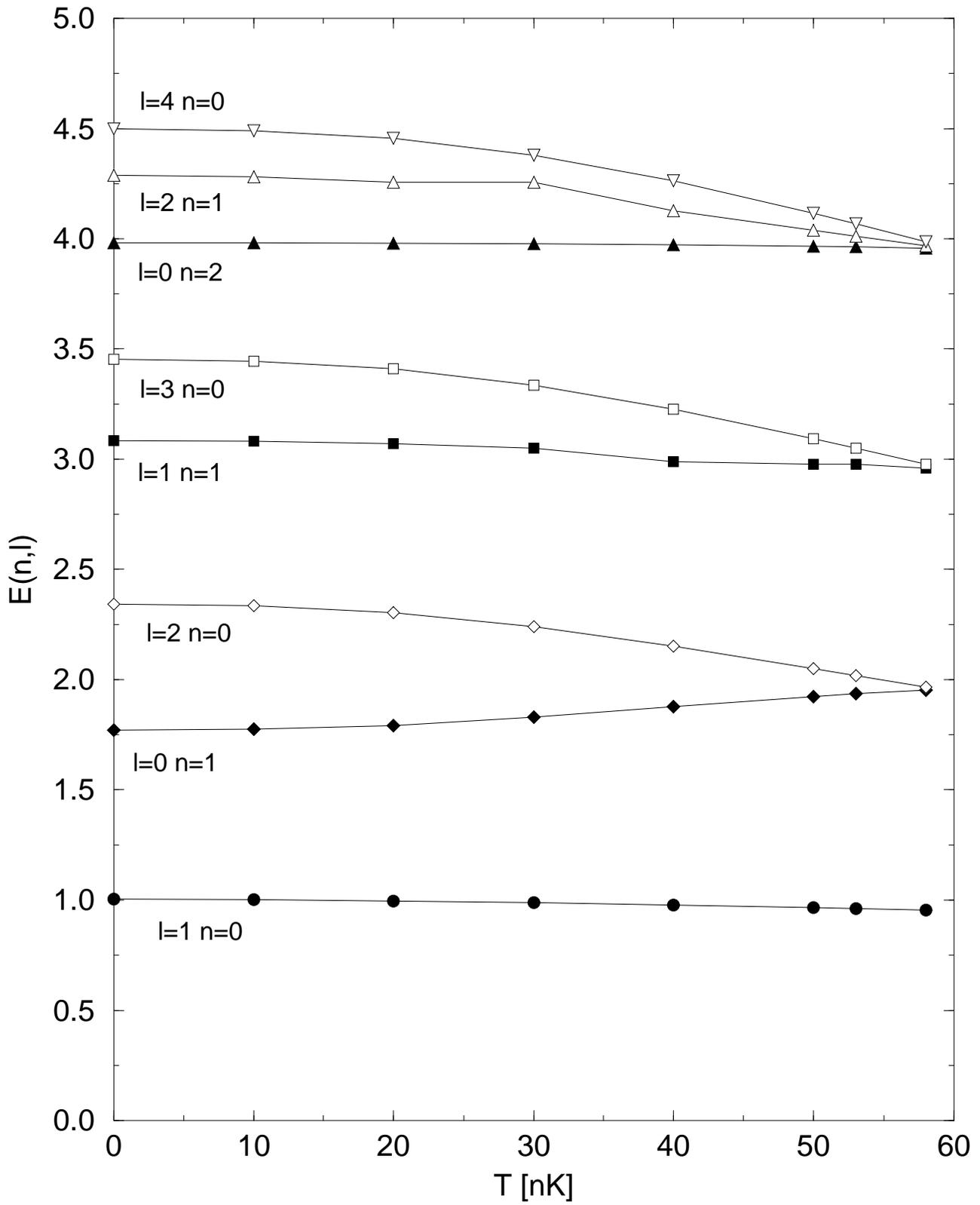
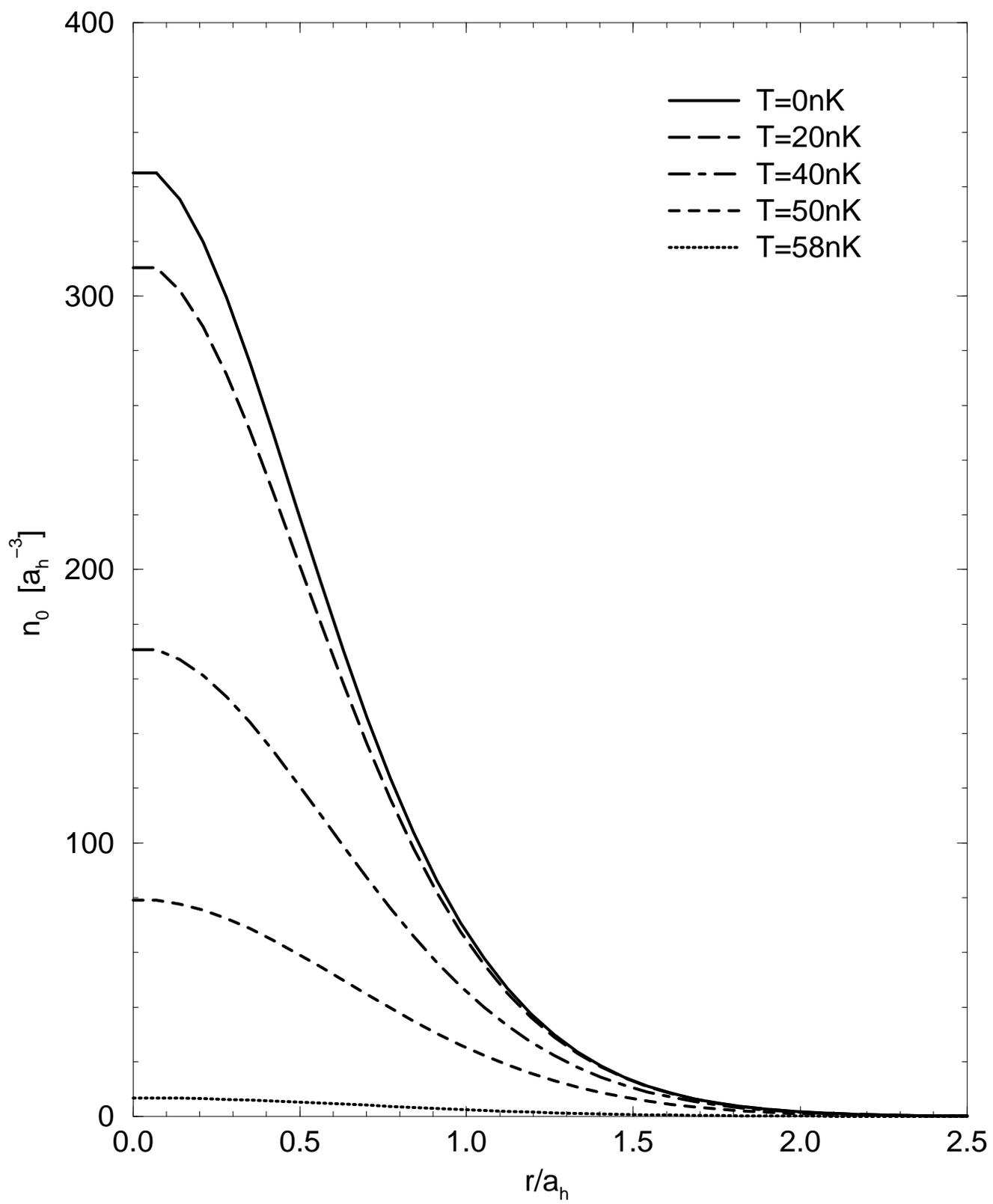
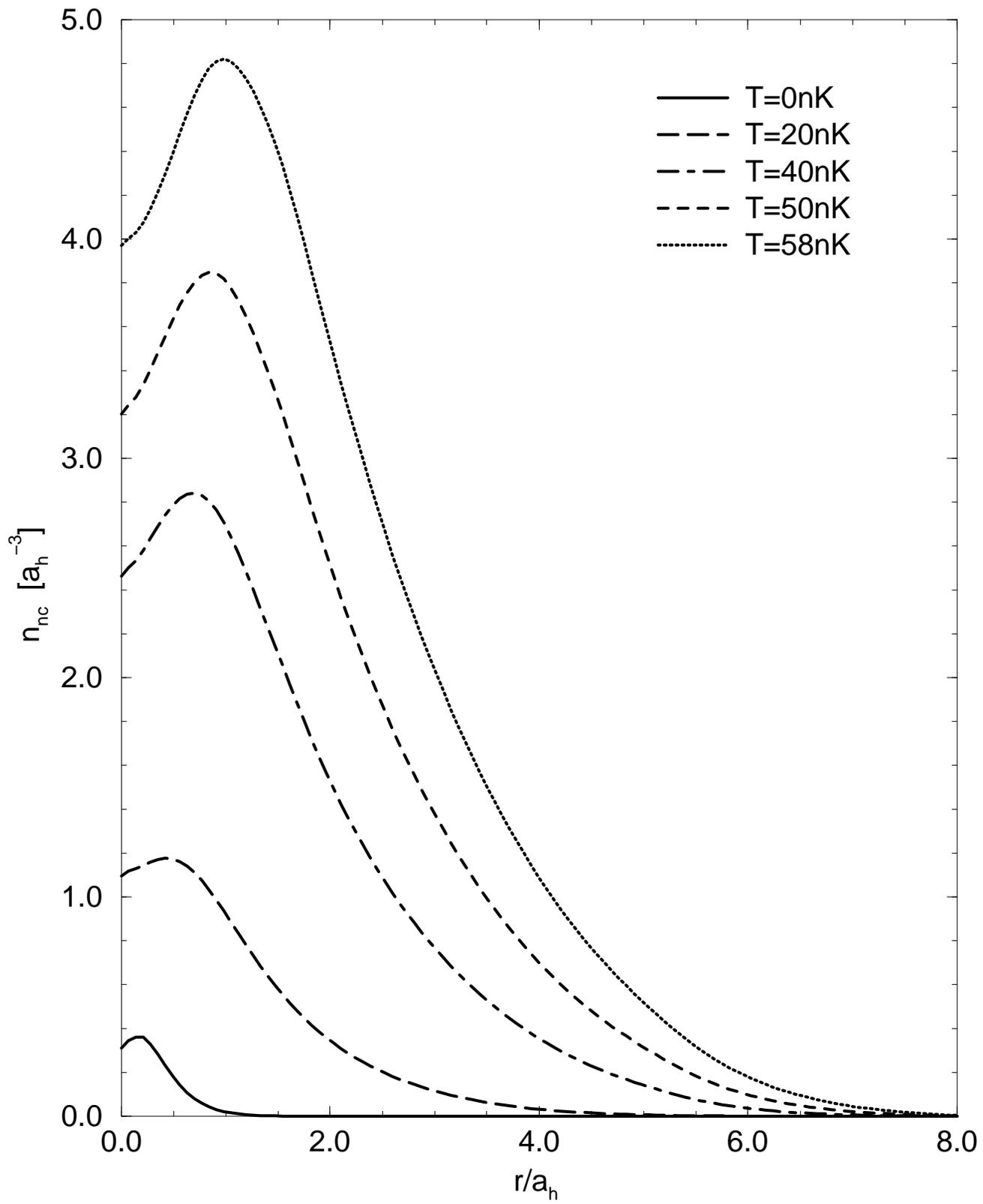
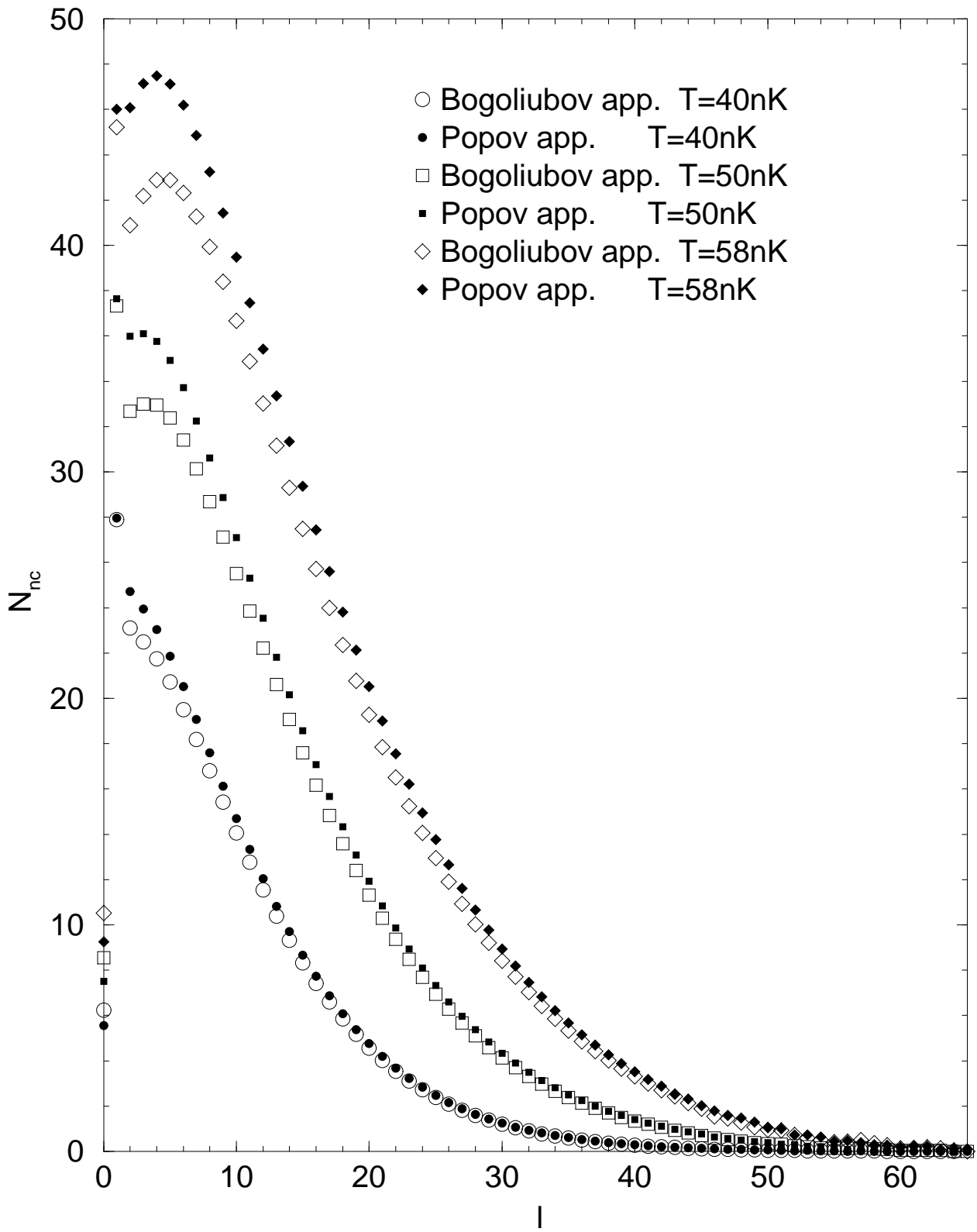
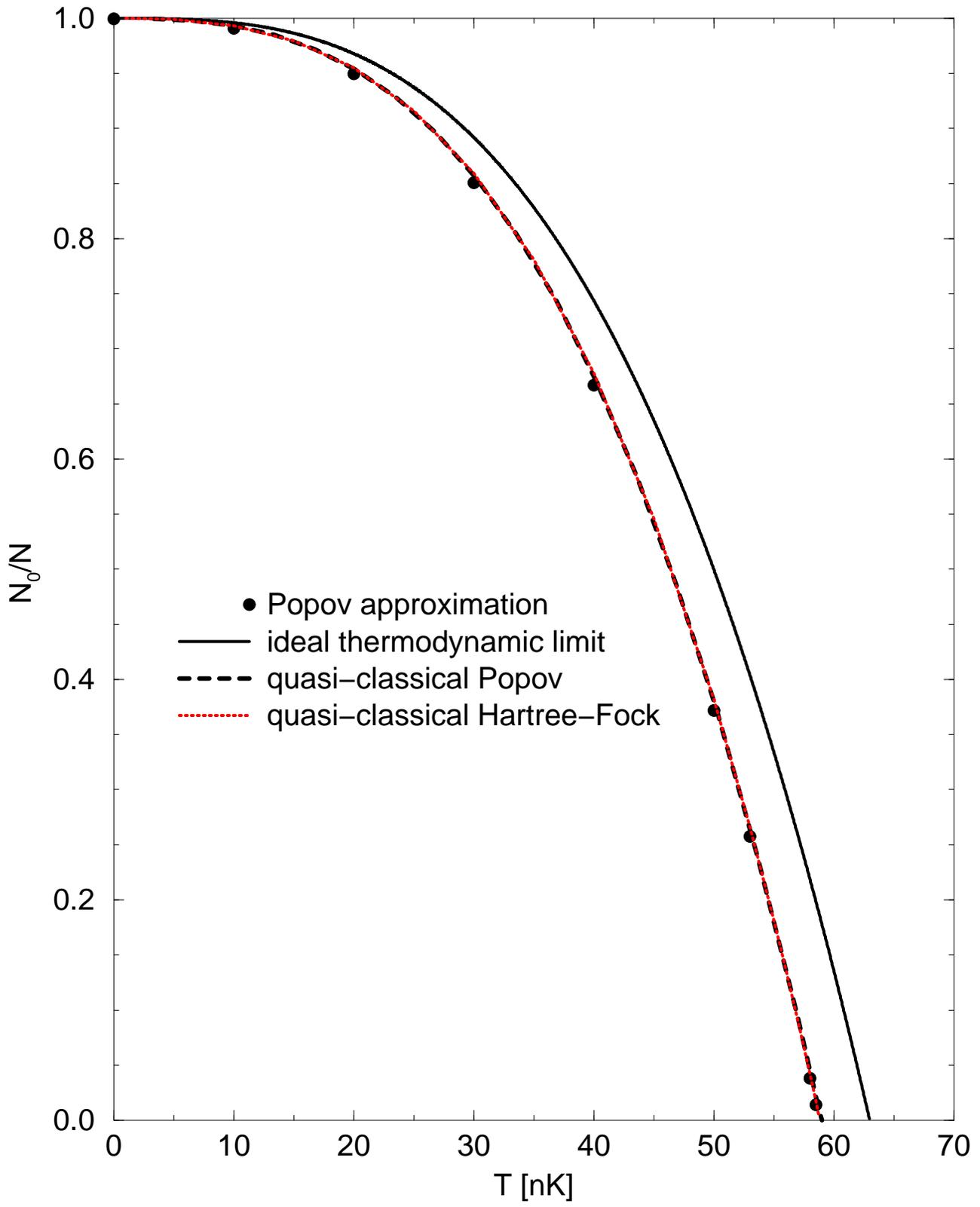
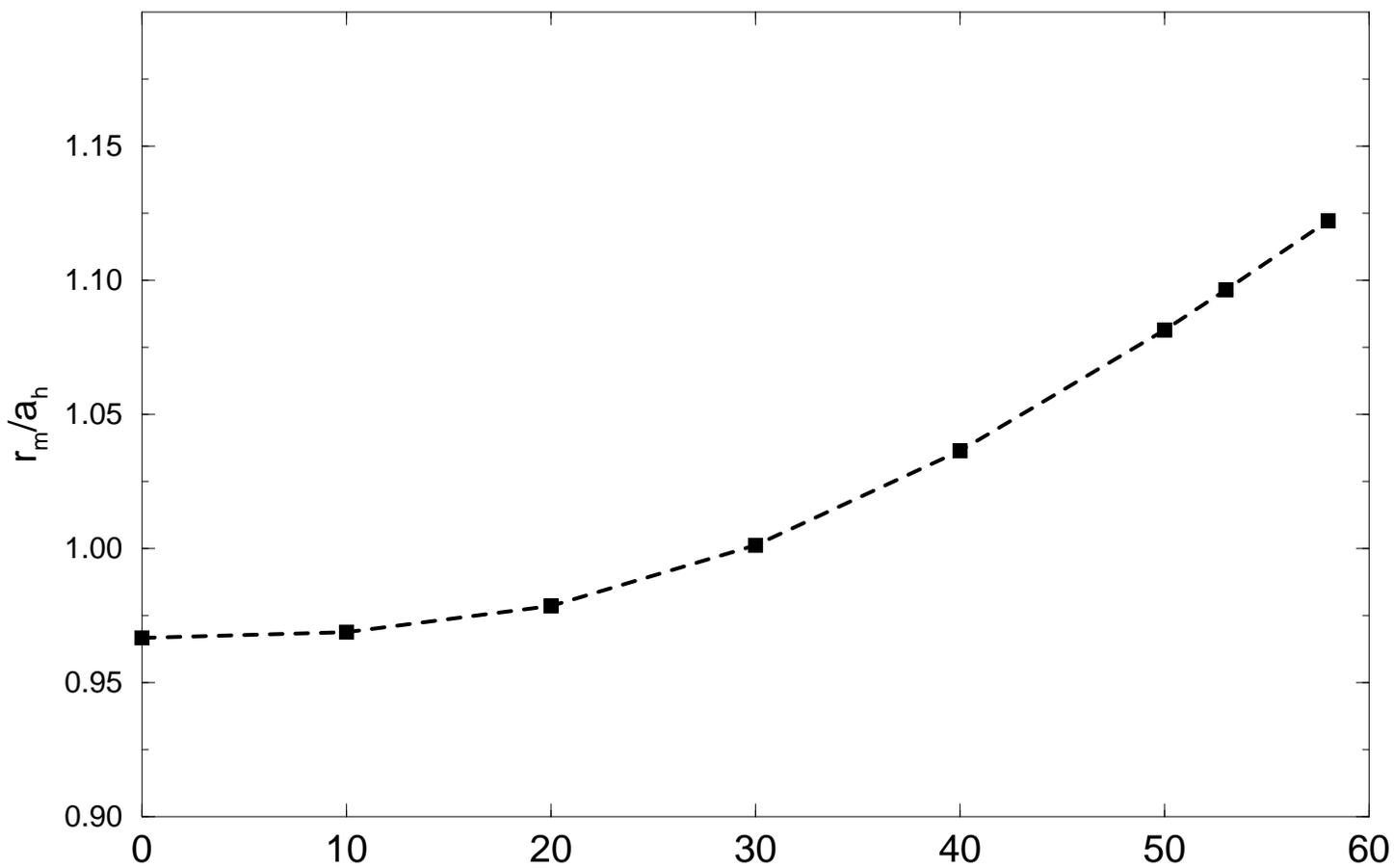
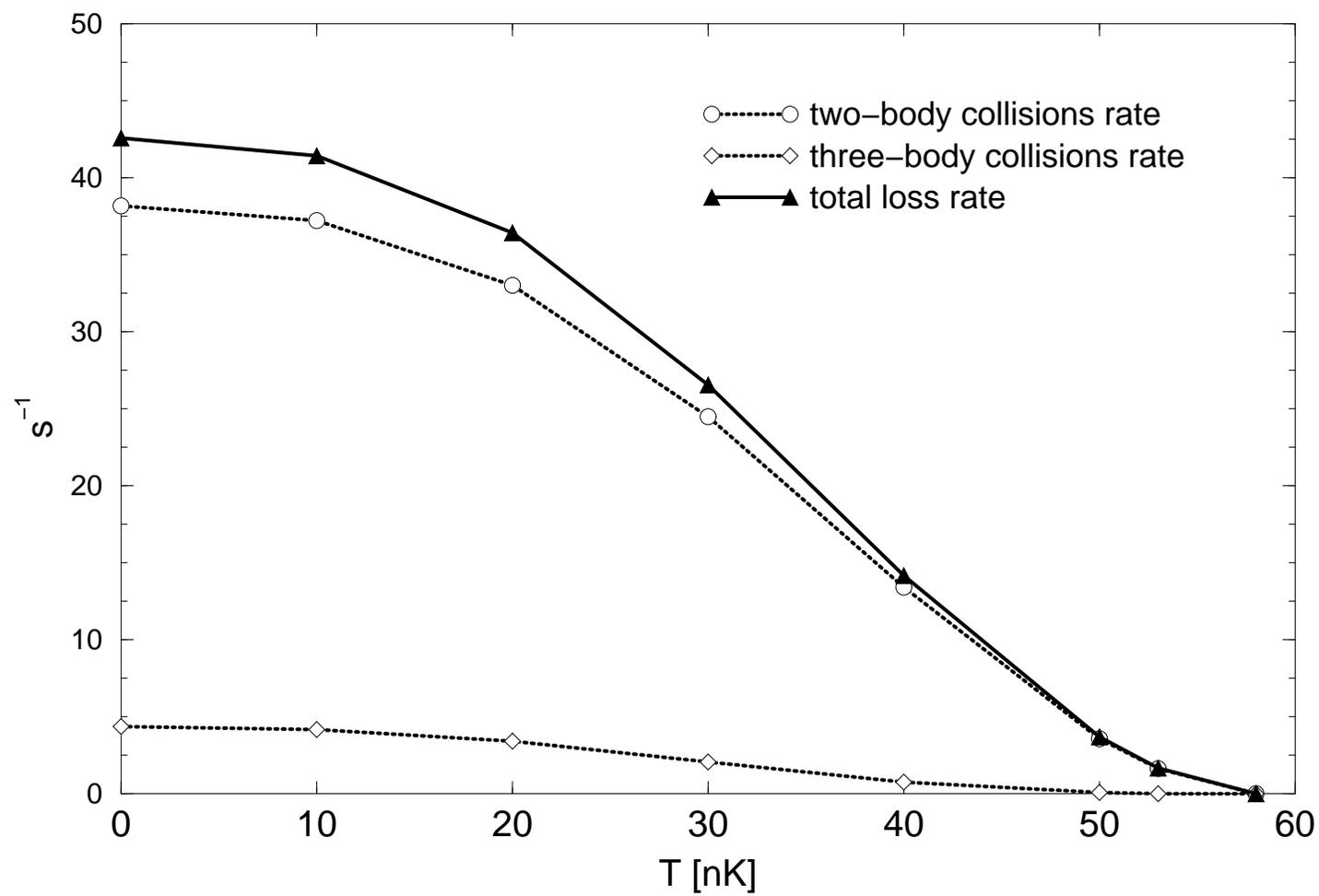
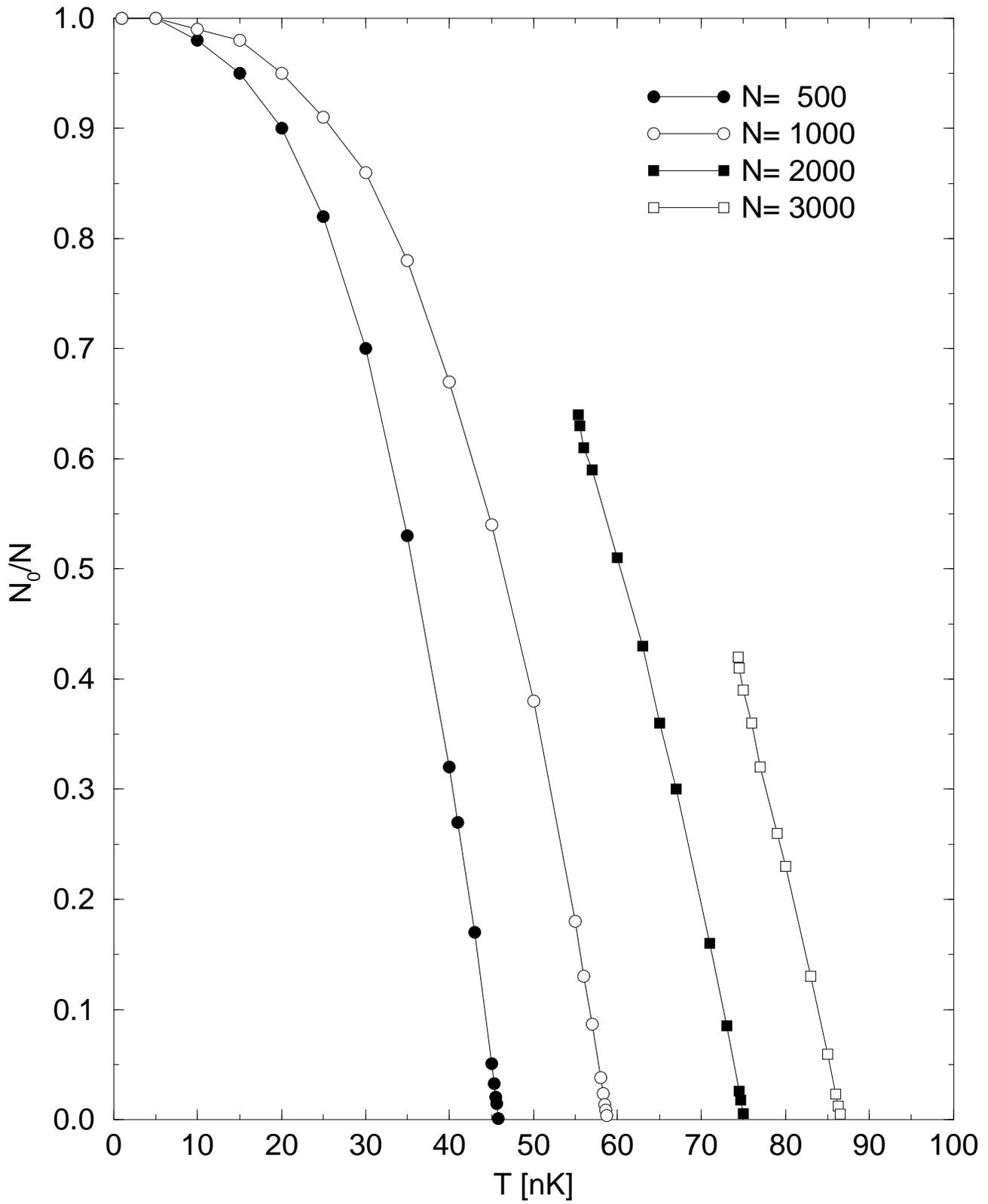
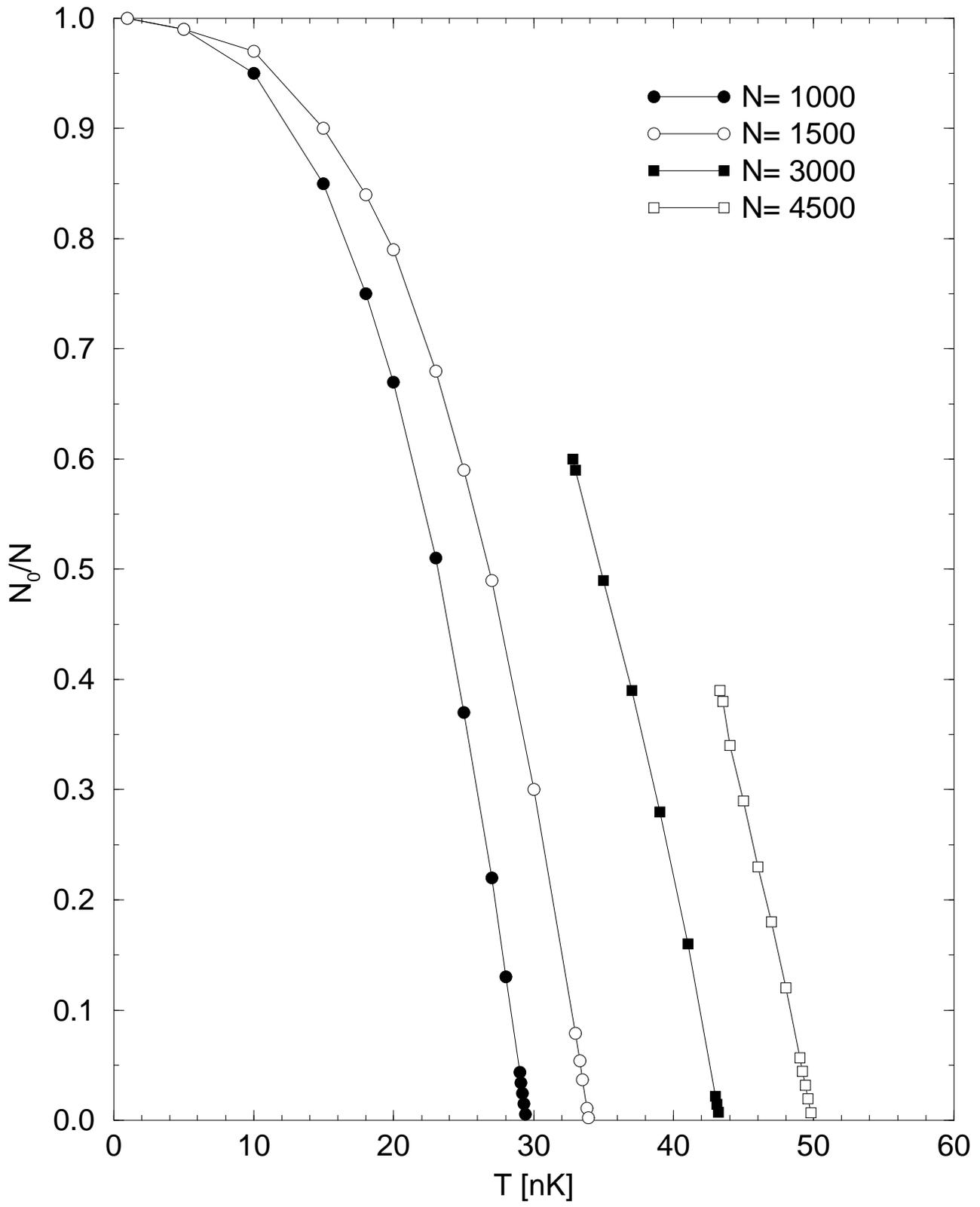
\end{document}